\documentstyle[12pt,aasms4]{article}
\lefthead{Cowley et al.}
\righthead{Magellanic Cloud X-ray Survey}

\begin{document}

\title{Magellanic Cloud X-ray Sources: III. Completion of a {\it ROSAT} Survey}
\author{P.C. Schmidtke\altaffilmark{1}, A.P. Cowley\altaffilmark{1}, \\
J.D. Crane\altaffilmark{2}, V.A. Taylor, and T.K. McGrath} 
\affil{Department of Physics \& Astronomy, Arizona State University, Tempe, AZ, 
85287-1504} 

\centerline{e-mail:  ~paul.schmidtke@asu.edu; ~anne.cowley@asu.edu}

\centerline{jdc2k@virginia.edu; ~taylor@antares.la.asu.edu; 
~mcgrath@sps.la.asu.edu}

\author{J.B. Hutchings\altaffilmark{1} and David Crampton\altaffilmark{1}} 
\affil{Dominion Astrophysical Observatory, National Research Council of
Canada, Victoria, B.C., \\ V8X 4M6, Canada} 

\centerline{e-mail: ~john.hutchings@hia.nrc.ca; ~david.crampton@hia.nrc.ca}

\authoremail{anne.cowley@asu.edu}

\altaffiltext{1} {Visiting Astronomers, Cerro Tololo Inter-American
Observatory, National Optical Astronomy Observatories, which is operated
by the Association of Universities for Research in Astronomy, Inc., under
contract with the National Science Foundation} 

\altaffiltext{2} {Now at Department of Astronomy, University of Virginia, 
Charlottesville, VA 22903-0818}
\begin{abstract}

This paper concludes a series of three papers presenting {\it ROSAT}
High-Resolution Imager (HRI) observations of unidentified {\it Einstein}
and serendipitous $ROSAT$ X-ray sources in the direction of the Magellanic
Clouds.  Accurate positions and fluxes have been measured for these
sources.  Optical photometry and spectroscopy were obtained to search for
identifications in order to determine the physical nature of these
sources.  The present paper includes new data for 24 objects;
identifications are given or confirmed for 30 sources.  For six sources
optical finding charts showing the X-ray positions are provided.  The
results from this program are summarized, showing the populations of
luminous X-ray sources in the Magellanic Clouds are quite different from
those in the Galaxy. 

\end{abstract}


\section{Introduction}

The first comprehensive X-ray survey of the Magellanic Clouds was carried
out by the {\it Einstein} Observatory which discovered a wide variety of
sources including many supernova remnants, a few luminous X-ray binaries,
and some foreground stars and background active galactic nuclei.  These
surveys were complete to about $L_X\sim3\times10^{35}$ erg s$^{-1}$ in
the energy range from 0.15 to 4 keV.  Catalogues of the compact and
point-like sources were published for the LMC (Long, Helfand, \&
Grabelsky 1981, Wang et al.\ 1991) and SMC (Seward \& Mitchell 1981, Wang
\& Wu 1992), and some optical identifications were made.  Cowley et al.\
(1984) and Crampton et al.\ (1985) extended the identification work and
found some additional counterparts, but many sources remained
unidentified, in part because of the relatively low positional accuracy
($\pm30^{\prime\prime}$) and the optically crowded fields in these nearby
galaxies. 

In an effort to study and compare the global populations of stellar X-ray
sources in the LMC, SMC, and the Galaxy, we undertook a program to obtain
accurate positions ($\pm5^{\prime\prime}$) for the unidentified point
sources using the $ROSAT$ High-Resolution Imager (HRI).  Two papers with
results from this survey have been published (Schmidtke et al.\ 1994,
hereafter Paper~I, and Cowley et al.\ 1997, hereafter Paper~II).  This
paper contains the third and final part of this study and a summary of
the results. 

\section{{\it ROSAT} X-ray Observations}

We present the new data obtained during the fifth year of the $ROSAT$
Guest Investigator program (AO5).  Table 1 lists the positions of the
centers of nineteen fields which were observed.  The exposure times are
also listed; their durations were chosen to provide at least a 5$\sigma$
detection of the faintest, unidentified {\it Einstein} source in the
field.  $ROSAT$ fields were not always centered on an individual source,
since sometimes several sources could be included in one image if the
centers were appropriately selected.  Dates of observation are listed
since these may prove to be useful in variability studies.  Some of the
X-ray observations were obtained in several segments, so multiple dates
are given. 

\subsection{Source Positions, X-ray Count Rates, and Upper Limits }

Tables 2 lists the point sources which were detected at a 3$\sigma$ level
or higher.  Both their $ROSAT$ and {\it Einstein} designations are given. 
The sources which have no {\it Einstein} name are serendipitous sources
detected in our $ROSAT$ images.  The count rates for all sources in this
table were measured in a 25$^{\prime\prime}$-diameter aperture, and the
background has been subtracted.  Since the image quality in the HRI
degrades off axis, we also list how far each source lies from the field
center.  Sources within 12$^{\prime}$ of the center should be well
measured both for position and count rate. 

In Table 3 we list sources which are clearly extended sources or for which
the count rate is slightly below a 3$\sigma$ detection, but where other
evidence indicates the source is real.  For example, RX~J0117.5$-$7312
falls below a 3$\sigma$ detection, but it is coincident with a relatively
bright foreground G star similar to others detected by $ROSAT$.  For most
of the extended sources it was necessary to use a larger aperture than our
standard 25$^{\prime\prime}$ diameter, so the fifth column lists the
aperture size used to measure the source.  The extended sources are
primarily supernova remnants (SNR) which were previously known. 

As detailed in Paper~I, the source positions were determined using PROS
software, but now it is unnecessary to apply a field rotation
(0.4$^{\circ}$) to correct for the satellite's pointing error (Kuerster
1993), since this adjustment is made by the standard pipeline data
processing.  Both Tables 2 and 3 give X-ray positions using equinox
J2000.0.  The point sources are estimated to have an uncertainty of
$\pm5^{\prime\prime}$, as was shown in Paper~II. 

\subsection{X-ray Variability }

Many of the sources are variable.  Table 4 lists the sources found by
{\it Einstein} but not detected by $ROSAT$.  Each of these sources should
have fallen within one or more of the $ROSAT$ fields, and its {\it
Einstein} count rate indicates it should have been detected by $ROSAT$. 
Thus, it is likely that many of these undetected sources are intrinsic
variables, although we cannot rule out the possibility that some were
spurious {\it Einstein} detections.  For example, from the {\it Einstein}
LMC data Wang et al.\ (1991) failed to find 26 sources originally
published by Long et al.\ (1981) from the same data.  A number of these
sources are also included in our ``not detected with $ROSAT$" lists.  Some
examples of sources not found in our $ROSAT$ images with a high level of
certainty (i.e. with a ratio of `expected' to upper-limit count rate $>5$)
are CAL 29, CAL 55, CAL 58, CAL 63, CAL 64, and CAL 96. 

In Table 4 we list the {\it Einstein} positions for our ``not detected"
sources, precessed to J2000.0, since we cannot measure their positions. 
In column 4 we give the 3$\sigma$ upper limit to the background-subtracted
count rate within a 100$^{\prime\prime}$ aperture, and in column 6 the
expected $ROSAT$ count rate, based on the {\it Einstein}-IPC count rate
and a blackbody spectral model.  Since the appropriate conversion factor 
depends on the unknown shape of each source's X-ray spectrum, a range of 
blackbody models has been assumed: $kT=0.3-1.0$ keV and a column density 
of log $N_H=21.4$ atoms cm$^{-1}$. 

For sources which were detected by both satellites, we can compare count
rates.  In Figure 1 we plot the observed $ROSAT$-HRI count rate versus
`expected' count rate based on the published {\it Einstein}-IPC flux. 
Upper limits for undetected sources are shown by arrows.  On average, the
`expected' count rates may be underestimated since we have assumed a hard
spectrum, and $ROSAT$ is very sensitive to low energy X-rays.  The plot
shows the sources detected by both instruments are quite variable (i.e.
they do not lie along a straight line).  Among the non-detected sources,
those which lie to the right of the dashed line are intrinsically
variable, while those to the left have not had a deep enough $ROSAT$
exposure to be observed. 

In all three papers we have noted in the ``$ROSAT$ X-ray Sources" tables
if a source was known to be variable.  However, this is very incomplete
since we only know about variability for cases where there are (a) several
observations, (b) the observed $ROSAT$ value is very discrepant from the
expected value, or (c) variability is noted in published literature.  It
was outside the scope of this study to follow individual sources for
variability.  Approximately 40\% of the sources are noted as variables,
but for the reasons given this is a lower limit to the true number of
variables. 

\section{Optical Observations}

\subsection{Photometry and Finding Charts}

During observing runs at CTIO in 1995 November and 1996 November, CCD
photometry was obtained using the 0.9-m telescope.  One or more $V$ frames
were taken of each source, and in many cases $UBV$ photometry was
obtained.  Both optical variability and color were used to select possible
candidates to be observed spectroscopically.  The magnitudes and colors
are based on DAOPHOT (Stetson 1987) analysis of the CCD images which have
been calibrated using observations of Landolt (1992) standard stars.  The
accuracy is $\sim0.02-0.03$ mag for the bright stars and up to 0.1 mag for
the very faintest stars. 

The $V$ magnitudes are given in Table 5 for all optically identified
sources that are not SNR.  The $V$ images were also used to prepare the
six finding charts shown in Figure 2.  Table 5 gives literature references
to published finding charts for the remaining sources.  Astrometry of each
CCD image was used to mark the X-ray positions and obtain the coordinates
of the optical counterparts given in Table 5.  The positions have been
tied to the coordinate system of the {\it HST Guide Star Catalog} (Lasker
et al.\ 1990) by measuring positions of 12 or more stars on the digitized
$GSC$ plate material. 

\subsection{Spectroscopy and Optical Identifications}

In 1996 November spectra were obtained of the best photometric candidates
using the CTIO 4-m telescope with the RC spectrograph and Reticon CCD. 
Two different gratings were used; the higher resolution covered the region
from 3900 -- 5020\AA\ at $\sim$0.9\AA\ per pixel, and the other grating
gave $\sim$1.8\AA\ per pixel in the region from 3720 -- 5850\AA.  A He-Ar
lamp was observed with each observation so that reliable velocities could
be measured.  

Table 5 lists the optically identified sources which lie in the fields
studied in this paper, although some of the counterparts were previously
known.  A few of sources from Paper~II are also included in Table 5 since
they were not identified at the time that manuscript went to press.  The
identifications with Magellanic Cloud members include: 4 Be stars, 1
massive X-ray binary (MXRB; LMC X-1), 1 planetary nebula, 1 supersoft
X-ray binary (CAL 87), 1 pulsar (PSR 0540$-$69), and 6 SNR.  Background
sources include 6 AGN, 1 galaxy, and 1 galaxy cluster.  Foreground objects
include 7 late-type galactic stars and a peculiar emission star (CAL 86).
Details about each of the sources are given in the following section. 

In Table 6 we summarize the identified sources from this survey which are
Magellanic Cloud members (including those from all three papers).  This, of
course, is not a complete listing of all known and optically identified
sources in the LMC and SMC.  We refer the reader to the catalogue by van
Paradijs (1995) and to the listing of supersoft sources in the Magellanic
Clouds by Kahabka \& van den Heuvel (1997) to supplement Table 6.  We
further note that sources continue to be identified (e.g. the recent
identification of additional Be stars by Buckley et al.\ 1998). 

\subsection{Comments on Individual Sources} 

In this section we give further information about each of the sources in
Tables 2 and 3.  The `CAL' number (sometimes referred to in the literature
as `LHG') is from Long et al.\ (1981), the `W' designation is from Wang et
al.\ (1991), and `SMC' numbers are from the listing by Wang \& Wu (1992). 

{\it RX~J0005.3$-$7427 = SMC 1:} This X-ray source is associated with a
Seyfert galaxy at redshift z=0.1316 which is slightly resolved on
ground-based images (Crampton et al.\ 1997).  A finding chart was given by
these authors. 

\vskip 10pt

{\it RX~J0033.3$-$6915 = SMC 70:} SMC 70 is very extended, with $ROSAT$
contours detected to a radius of $\sim2^{\prime}$ (see finding chart in
Crampton et al.\ 1997).  It appears to be associated with the galaxy
cluster Abell 2789.  The X-ray source is centered on the central cD
galaxy whose redshift is z=0.0975. 

\vskip 10pt

{\it RX~J0054.9$-$7227 = SMC 35:} SMC 35 was observed far off axis
(18.8$^{\prime}$), and hence its X-ray position is poor.  A finding chart
for the field is given in Fig.\ 2.  The X-ray source, also designated as
XTE J0055$-$724 and ISAX J0054.9$-$7226, has been found to be a 59-s X-ray
pulsar (Marshall \& Lochner 1998).  Thus, Star 1 which is a Be star
($V=16.6$, $B-V=-0.05$, $U-B=-0.78$) is a possible candiate as the neutron
star's companion.  However, its position at RA=$00^h54^m51.1^s$,
Dec=$-72^{\circ}27^{\prime}04.0^{\prime\prime}$ (2000.0) is
31.5$^{\prime\prime}$ away from the X-ray position.  Stars marked as 2, 3,
and 4 we have measured magnitudes and colors, but no spectra have been
obtained.  Star 2 has $V=15.23$, $B-V=-0.03$, $U-B=-0.92$; Star 2 has
$V=16.28$, $B-V=-0.11$, $U-B=-0.70$; Star 3 has $V=15.59$, $B-V=-0.04$,
$U-B=-0.46$.  These stars are possible alternative candidates. 
   
\vskip 10pt

{\it RX~J0058.2$-$7231:} This very weak source from Table 3 has a count
rate which is slightly below the 3$\sigma$ level (0.0031$\pm$0.0011).
However, it is located only 5$^{\prime\prime}$ from a 15$^{th}$ magnitude
Be star which is likely to be the optical counterpart.  A finding chart
for this source is given in Fig.\ 2 where the Be star is marked. 

\vskip 10pt

{\it RX~J0058.6$-$7136 = SMC 43 = 1E0056.8$-$7154:} SMC 43 was identified
with the SMC planetary nebula N67 by Wang (1991) on the basis of an {\it
Einstein}-HRI position.  The source is discussed in Paper~II, and a finding
chart is given there (`Star' 2).  SMC 43 is included in this listing
because a new $ROSAT$ observation provided a slightly improved X-ray
position.  This source is especially interesting since its energy spectrum
shows it to be one of rare class of supersoft X-ray sources. 

\vskip 10pt

{\it RX~J0059.4$-$7210 = SMC 44:} SMC 44 is a very extended source (note
100$^{\prime\prime}$ aperture used for detection).  It is associated with
the supernova remnant N66 (= NGC 346 = SNR 0057$-$724 = DEM S103) and is
claimed to be the largest and most luminous H~II region in the SMC (Ye,
Turtle, \& Kennicutt 1991).  No finding chart is given here. 

\vskip 10pt

{\it RX~J0100.7$-$7212 = SMC 45:} SMC 45 lies quite far from the center of
the $ROSAT$ field (13.7$^{\prime}$).  It is a weak point source.  These
facts suggest its position has rather low accuracy.  We have identified it
with a foreground G star ($V=13.5$) which lies 13$^{\prime\prime}$ from
the X-ray position.  A finding chart is given in Fig.\ 2. 

\vskip 10pt

{\it RX~J0117.5$-$7312:} The X-ray observations for this source are from a
field (Observation \#28) observed for Paper~II.  However, the data were
not complete at the time that paper was published, so the source is
included here.  It is identified with a foreground G star ($V=13.9$).  The
positional coincidence is good, as can be seen in the finding chart in
Fig.\ 2. 

\vskip 10pt

{\it RX~J0136.4$-$7105 = SMC 68:} This point X-ray source is nearly
coincident with a Seyfert galaxy ($V=19.1$) whose redshift is z=0.4598. 
Its spectrum and finding chart are shown by Crampton et al.\ (1997). 

\vskip 10pt

{\it RX~J0454.2$-$6643:} This source has also been identified by Crampton
et al.\ (1997) with a Seyfert galaxy ($V=18.2$) at redshift z=0.2279. 
These authors also show the spectrum and give an optical finding chart. 

\vskip 10pt

{\it RX~J0518.8$-$6939 = CAL 23:} The X-ray contours for this source are
very extended.  Its optical/radio counterpart is the SNR N120 (=
0519$-$697).  No finding chart is given in this paper. 

\vskip 10pt

{\it RX~J0519.8$-$6926 = CAL 27 = W520.1$-$6929:} CAL 27 is another very
extended supernova remnant (SNR 0520$-$694).  It was measured and reported
on in Paper~I.  The present data are from a second $ROSAT$ image.  We give 
no finding chart. 

\vskip 10pt

{\it RX~J0531.5$-$7130 = CAL 46 = W532.3$-$7132:}  CAL 46 was observed in
Paper~I and identified with a Seyfert galaxy (`Star' 1) whose redshift is
z=0.2214.  The new $ROSAT$ observations reported here show that the source
is variable, as its count rate decreased by a factor of two. 

\vskip 10pt

{\it RX~J0534.0$-$7145:} This point X-ray source lies near but not
directly in the center of the large S0 galaxy, Up 053448$-$7147.3 (see
optical image and marked X-ray position on the finding chart given by
Crampton et al.\ 1997).  The optical spectrum shows narrow [O~II] and
[O~III] emission lines, but otherwise it appears relatively normal.  The
galaxy's redshift is z=0.0242. 

\vskip 10pt

{\it RX~J0536.0$-$6736 = CAL 60 = W536.2$-$6737:} This very extended
source was also observed in another $ROSAT$ image as reported in Paper~I. 
It is identified with the SNR 0536$-$676 (Mathewson et al.\ 1985) = DEM
241 (Davies et al.\ 1976).  However, Wang et al.\ (1991) found the source
to be variable.  There may be one or more individual sources blended with
the supernova remnant.  A finding chart was given in Paper~I. 

\vskip 10pt

{\it RX~J0537.8$-$6910 = CAL 67 = W538.1$-$6912 = NGC 2060:} CAL 67 lies
near the center of the complex 30 Doradus region.  It is associated with
the supernova remnant N157B (Henize 1956), also known as SNR 0538$-$691,
DEM 263, or 30 Dor B.  It is a Crab-like SNR (Wang \& Gotthelf 1998)
containing a recently discovered 16 ms pulsar (Marshall et al.\ 1998).
Data for this source were described in Paper~II based on measurements from
a different $ROSAT$ image. 

\vskip 10pt

{\it RX~J0538.3$-$6924 = CAL 69 = W538.7$-$6925:} Although CAL 69 was
identified in Paper~II with a galactic dMe star, its X-ray position and
count rate were both very questionable since the $ROSAT$ source was
observed 13$^{\prime}$ off axis.  A new $ROSAT$ image, with the source
centered in the field, confirms the identification and provides the count
rate given in Table 2.  The optical identification is with Star 1, as
shown in the finding chart in Paper~II. 

\vskip 10pt

{\it RX~J0538.6$-$6853 = CAL 71 = W538.9$-$6855:} Like CAL 69, this source
was far from the field center in the Paper~II observation, so its position
and count rate were suspect.  The new $ROSAT$ data show that it is
correctly identified with the RS CVn star marked in the finding chart in
Paper~II.  It is unclear whether the source is variable (count rates were
0.019$\pm$0.002 cts s$^{-1}$ in Paper~II versus 0.053$\pm$0.005 cts
s$^{-1}$ in the present paper) or if the Paper~II value was extremely low
because the source was observed so far off axis ($\sim16^{\prime}$). 

\vskip 10pt

{\it RX~J0538.8$-$6910 = CAL 73 = W539.2$-$6911:} This extended source
required a very large aperture for detection (300$^{\prime\prime}$).  It
lies in the 30 Dor complex and may possibly be a supernova remnant, but
this is not certain.  It certainly is part of the hot gas in this region
of the LMC.  This region, including all of 30 Dor, has been mapped with
deeper $ROSAT$ observations by Wang (1995). 

\vskip 10pt

{\it RX~J0540.2$-$6920 = CAL 79 = PSR 0540$-$69:}  This very strong source
is the well-known pulsar PSR 0540$-$69.  We show its optical image in
Fig.\ 2.  Our integrated magnitude of the plerion within a
12$^{\prime\prime}$ aperture is $V=17.5$.  The position and an extensive
discussion are given by Caraveo et al.\ (1992) and by Shearer et al.\
(1994). 

\vskip 10pt

{\it RX~J0546.3$-$6835 = CAL 86? = W546.6$-$6836:} Although this source
was measured in Paper~I and a finding chart was given, it was not
identified at that time.  The X-ray source may be variable, although the
value in Paper~I was measured from a very off-axis position.  The optical
counterpart is a variable star, Star 2 ($V=19.5_{var}$).  The spectrum
shows double hydrogen and He I emission lines and is variable.  No period
has yet been determined, but it appears to be only a few hours.  It is
clearly related to some type of cataclysmic variable and lies in the
Galaxy. 

\vskip 10pt

{\it RX~J0546.8$-$7109 = CAL 87:} CAL 87 is a well-known, eclipsing,
supersoft X-ray binary (see Cowley et al.\ 1990, Hutchings et al.\ 1998,
and references therein).  A finding chart is given by Schmidtke et al.\
(1993). 

\vskip 10pt

{\it RX~J0549.8$-$7150:}  The finding chart in Fig.\ 2 shows that the
source is identified with a $V=13.5$ galactic dMe star.  The optical
spectrum shows both hydrogen and Ca~II in emission.  This source was
blended with a nearby transient supersoft source (RX~J0050.0$-$7151) in
the original $ROSAT$-PSPC image because it was observed far off axis where
the point-spread function is extremely large.  Subsequent analysis of the
All-Sky Survey data by Thomas (1996) has resolved these two sources.  
RX~J0549.8$-$7150 is the only source present in our HRI data, as the
supersoft source had faded below the detection level.  The identification
of this source has been discussed by Schmidtke \& Cowley (1996) and
Charles et al.\ (1996). 

\vskip 10pt

{\it RX~J0550.5$-$7110:} This very weak point source has been identified
with a Seyfert galaxy ($V=19.9$) whose redshift is z=0.4429.  Its spectrum
and finding chart are shown by Crampton et al.\ (1997). 

\vskip 10pt

\subsection{Sources Which Remain Unidentified}

Almost all of the sources which were detected with $ROSAT$ in this study
have been identified.  Of the previously known sources, there are only
five (from all three papers) for which we have not found optical
counterparts.  Surprisingly, they are not all weak, and hence uncertain,
sources.  Two of them have strong count rates: W547.4$-$6853
(RX~J0547$-$6852) has a count rate of 0.0174$\pm0.0038$ cts s$^{-1}$ and
CAL 93 (RX~J0552.5$-$6949) was observed at 0.0051$\pm$0.0013 cts s$^{-1}$.
The other three previously known but still unidentified sources are
weaker; they are W517.4$-$7149 (RX~J0516.7$-$7146; 0.0032$\pm$0.0010 cts
s$^{-1}$), W534.9$-$6741 (RX~J0534$-$6739; 0.0045$\pm$0.0011), and
W536.7$-$7043 (RX~J0536.0$-$7041; 0.0033$\pm$0.0011).  Of the
serendipitous $ROSAT$ sources in this survey, we have been unable to find
optical counterparts for RX~J0515.1$-$7144, RX~J0527.8$-$6954,
RX~J0551.4$-$7028, and RX~J0552.8$-$6927 (which may be the same as CAL 94,
but the positional agreement is poor).  We note that RX~J0527.8$-$6954 is
a transient supersoft source which has been searched for by a number of
workers (e.g. Cowley et al.\ 1993, Greiner et al.\ 1996).  Alcock et al.\
(1997) have suggested the counterpart may be HV 2554, but the
identification is uncertain. 

\section{Conclusion and Summary}

The goal of this project was to compare the stellar X-ray source
population in the Magellanic Clouds with that in the Galaxy.  After
removing the X-ray-bright foreground stars, background galaxies, and SNRs
from the sample, one is left with a fairly small population of X-ray
sources which are Cloud members.  They are spread over a wide area in each
galaxy, with a radius of $\sim8^{\circ}$ in the LMC and $\sim3^{\circ}$ in
the SMC.  In the LMC they are not concentrated to the star-rich Bar. 

In spite of the fact that quite a number of discrete {\it Einstein}
sources were not recovered in this {\it ROSAT} survey, we still can make
an interesting intercomparison between the types of X-ray sources with
$L_X\geq3\times10^{35}$ erg s$^{-1}$.  Data on galactic X-ray sources are
taken from the catalogue by van Paradijs (1995).  In the Galaxy one finds
that there are $\sim$50 massive X-ray binaries (MXRB), divided
approximately equally between those with early-type supergiant primaries
and the longer-period systems containing Be stars.  Since the LMC contains
about one-tenth the mass of the Galaxy, one expects to find only a small
number of MXRB.  However, the LMC contains 4 MXRB with supergiant
primaries and at least a dozen Be-star systems.  In the SMC there is only
one supergiant MXRB (SMC X-1), but this galaxy also contains about a dozen
Be-type MXRB.  The number of X-ray bright Be stars is still increasing in
both galaxies with new data from $ASCA$ and $RXTE$.  Thus, both of the
Magellanic Clouds are surprisingly rich in massive early-type X-ray
binaries containing Be stars.  Since Be stars are rapidly rotating, this
may point to a difference in rotational properties of early-type stars in
the Galaxy and the Magellanic Clouds. 

The story is quite different for the low-mass X-ray binaries (LMXB).  LMXB
are very numerous in the Galaxy, with approximately 150 being known in the
field and 12 in globular clusters.  They include some X-ray transients,
bursting sources, X-ray pulsars, and persistent sources.  In the Galaxy
the LMXB are known to be part of an old stellar population based on their
kinematics, chemical compositions, and spatial distribution.  By contrast,
the \underbar{only} LMXB known in the Magellanic Clouds is LMC X-2.  None
is found in the SMC or in globular clusters of these galaxies.  Thus, in
Magellanic Clouds (considered together) the ratio MXRB/LMXB $\sim25$ while
in the Galaxy this ratio is $\sim0.3$.  This marked difference in the
X-ray source populations is shown as a histogram by Cowley et al.\ (1998,
see their Fig.\ 1).  We emphasize that our X-ray detection limit in the
Clouds is well below the luminosity of all the galactic X-ray binaries,
both high and low mass. 

Among the supersoft X-ray binaries, seven are known in the LMC and four in
the SMC, but only two in the Galaxy (Kahabka \& van den Heuvel 1997).
However, this difference appears to be a selection effect rather than a
true population indicator, since in the Galaxy these sources are highly
absorbed by interstellar hydrogen gas in the plane.  The lower column
density of gas towards the LMC and SMC makes them more easily detected
there. 

In summary, our survey of previously unidentifed {\it Einstein} point
sources in the direction of the Magellanic Clouds has been completed.  We
have found the optical counterparts for nearly all of the sources which
were recovered.  However, because a high percentage of sources are
variable, some were not detected in our survey.  Previously unknown
sources were also found for the same reason.  After removal of the
foreground and background sources from the sample, only a small number of
LMC and SMC members remain.  These show distinct differences compared to
the Galaxy, with the oldest stellar population almost entirely missing
from the Magellanic Clouds. 

Much deeper $ROSAT$ images of parts of the Magellanic Clouds are available
in the archives.  Systematic deep surveys of the LMC and SMC are underway
by Chu, Snowden, and collaborators (Chu et al.\ 1996) which reach an order
of magnitude fainter.  We are in the process of studying these data.  We
have created a preliminary point-source catalogue, and optical
observations are underway.  These data should add further to our
understanding of the X-ray source populations in the Magellanic Clouds. 

\acknowledgments

This work has been supported by both NSF and NASA.  

\clearpage

\begin{table}
\caption[]{Observed {\it ROSAT} Fields in SMC \& LMC }
\begin{flushleft}
\begin{tabular}{cllc}
\hline
\hline
Obs. & ~~R.A. ~~~~~~~~Dec. & Date(s) of & Exp. Time \\
\# &~~~~~~(J2000.0) &Observation & (ksec) \\
\hline
50 & 00$^h$05.5$^m$  $-$74$^\circ$ 27$^\prime$ & 1994 Oct 30 & 1.32 \\
51   & 00~~33.6 ~ $-$69~~16 & 1995 May 14             & 2.10 \\
52   & 00~~58.6 ~ $-$71~~36 & 1995 May 14             & 1.95 \\
53   & 00~~59.0 ~ $-$72~~23 & 1995 Apr 14, 18         & 5.00 \\
54   & 01~~36.4 ~ $-$71~~05 & 1994 Dec 15; 1995 May 15  & 3.21 \\
55   & 04~~54.2 ~ $-$66~~43 & 1995 May 26, 28         & 2.59 \\
56   & 04~~58.9 ~ $-$66~~28 & 1995 Feb 14, Mar 4      & 2.15 \\
57   & 05~~20.5 ~ $-$69~~32 & 1995 Jul 22, 27         & 3.83 \\
58   & 05~~27.8 ~ $-$69~~54 & 1995 Jul 30, Sep 27     & 4.87 \\
59   & 05~~32.0 ~ $-$69~~20 & 1995 Apr 5; Sep 19, 20  & 5.89 \\
60   & 05~~32.3 ~ $-$71~~40 & 1995 Aug 23, 24         & 5.31 \\
61   & 05~~36.3 ~ $-$67~~18 & 1995 Jul 9, 17          & 4.58 \\
62   & 05~~38.3 ~ $-$69~~24 & 1995 Aug 12, 13         & 2.45 \\
63   & 05~~38.9 ~ $-$69~~02 & 1995 Apr 7, 9           & 2.74 \\
64   & 05~~44.1 ~ $-$71~~01 & 1995 Jul 30, 31         & 6.00 \\
65   & 05~~47.9 ~ $-$68~~23 & 1995 Jul 2, 17          & 5.30 \\
66   & 05~~48.5 ~ $-$71~~13 & 1995 Sep 18, 19         & 9.39 \\
67   & 05~~50.0 ~ $-$71~~51 & 1995 Sep 7, 10          & 2.17 \\
68   & 05~~53.2 ~ $-$67~~54 & 1995 Jul 22, 28         & 7.80 \\
\hline
\end{tabular}
\end{flushleft}

\end{table}

\clearpage

\begin{scriptsize}

\begin{table}
\caption[]{SMC \& LMC X-Ray Point Sources}
%\begin{flushleft}
\begin{tabular}{llcccl}
\hline
\hline
Source$^a$ & ~~~~R.A. ~~~~~~~~~~~~ Dec. &Count Rate$^b$  
& Obs.$^c$ &~~Offset$^d$ & Notes$^e$ \\
RX~J &~~~~~~~~~~(J2000.0) & (cts s$^{-1}$ $\pm$1$\sigma$) & \# &( $\prime$ )\\  
\hline
0005.3$-$7427  & 00$^h$05$^m$19.5$^s$  $-$74$^\circ$ 26$^\prime$ 
42$^\prime$$^\prime$ &0.0315$\pm$0.0058 & ~50 & 0.7  \\
 ~~~~~~= SMC 1 \\
0058.6$-$7136 & 00~~58~~37.3 ~~$-$71~~35~~49 &0.0545$\pm$0.0058 &52 &0.2 & 
1. \\   
~~~~~~= SMC 43 &= 1E0056.8$-$7154 \\
0100.7$-$7212 & 01~~00~~42.4 ~~$-$72~~11~~30 &0.0030$\pm$0.0010 & 53 & 13.7 \\ 
 ~~~~~~= SMC 45 \\ 
0136.4$-$7105 & 01~~36~~25.6 ~~$-$71~~05~~25 &0.0078$\pm$0.0020 &54 & 0.6 & \\ 
 ~~~~~~= SMC 68 \\
0454.2$-$6643  & 04~~54~~10.6 ~~$-$66~~43~~17 &0.0082$\pm$0.0023 &55 &0.2 \\

0531.5$-$7130 & 05~~31~~31.9 ~~$-$71~~29~~44 &0.0069$\pm$0.0014 &60 & 11.1 & 
P1; var \\  
~~~~~~= CAL 46 &= W532.3$-$7132 \\ 
0534.0$-$7145 & 05~~34~~00.5 ~~$-$71~~45~~28 &0.0124$\pm$0.0018 & 60 &9.7 \\ 
0538.3$-$6924 & 05~~38~~16.4 ~~$-$69~~23~~30 &0.0493$\pm$0.0049 & 62 &0.5 & 
P2; var? \\
 ~~~~~~= CAL 69 &= W538.7$-$6925\\
0538.6$-$6853 &  05~~38~~34.4 ~~$-$68~~53~~03 &0.0531$\pm$0.0048 &63 &8.9 & 
P2; var? \\ 
 ~~~~~~= CAL 71 &= W538.9$-$6855 \\
 0546.3$-$6835 & 05~~46~~15.5 ~~$-$68~~35~~26 &0.0030$\pm$0.0010 &65 &15.5 & 
P1; var \\ 
 ~~~~~~= CAL 86? &= W546.6$-$6836 \\
0546.8$-$7109 &  05~~46~~47.2 ~~$-$71~~08~~53 &0.0501$\pm$0.0024 & 66 & 9.4 & 
2.; var \\
 ~~~~~~= CAL 87 \\ 
0549.8$-$7150 & 05~~49~~46.7 ~~$-$71~~49~~38 &0.0084$\pm$0.0025 & 67 & 1.7 \\ 
0550.5$-$7110 & 05~~50~~32.1 ~~$-$71~~09~~56 &0.0024$\pm$0.0007 & 66 &
10.3 \\
\hline
\\
\end{tabular}

$^a$~SMC sources (Wang \& Wu 1992);
CAL$=$LHG sources (Long et al.\ 1981); W sources (Wang et al.\ 1991) \\
$^b$~All values are background subtracted count rates using
 25$^\prime$$^\prime$-diameter aperture \\
$^c$~From Table 1 \\
$^d$~Angular distance of source from center of field \\
$^e$~P1 or P2 indicates source was measured in Paper I or II in
another $ROSAT$ image \\
{\bf Notes:} \\
1.~~Planetary nebula and supersoft X-ray source (see Table 5) \\
2.~~Also in Obs.\#64 at 0.0160 $\pm0.0018$ ct s$^{-1}$ and
15.1$^{\prime}$ from center; eclipsing supersoft binary (see Table 5) \\

%\end{flushleft}

\end{table}  
\end{scriptsize}

\clearpage

\begin{scriptsize}

\begin{table}
\caption[]{Extended or Weak SMC \& LMC X-Ray Sources}
%\begin{flushleft}
\begin{tabular}{llccccl}
\hline
\hline
Source$^a$  &~~~R.A. ~~~~~~~~~~~~ Dec. &Count Rate   
&Aperture &Obs$^b$ & Offset$^c$ & Notes$^d$ \\ 
RX~J &  ~~~~~~~~~(J2000.0) & (cts s$^{-1}$ $\pm$1$\sigma$) & 
( $\prime\prime$ ) &\#  &( $\prime$ ) \\
\hline
0033.3$-$6915 & 00$^h$33$^m$20.1$^s$ 
$-$69$^\circ$15$^\prime$ 14$^{\prime\prime}$ 
&0.0139$\pm$0.0032 & 30 & 51 & 1.7 \\
~~~~~~~= SMC 70 \\
0054.9$-$7227  & 00~~54~~56.3 ~~$-$72~~26~~43 &0.0077$\pm$0.0019 & 100 &53 
&18.8 & 1.; var; pulsar  \\ 
~~~~~~~= SMC 35 & & & & & & \\ 
0058.2$-$7231 & 00~~58~~12.7 ~~$-$72~~30~~45 &0.0031$\pm$0.0011 & 30 &53 
&8.7 \\ 
0059.4$-$7210  & 00~~59~~25.0 ~~$-$72~~10~~03 &0.0158$\pm$0.0025 & 
100 & 53 & 12.9 &SNR \\  
~~~~~~~= SMC 44 \\
0117.5$-$7312 & 01~~17~~29.4 ~~$-$73~~11~~38 &0.0018$\pm$0.0008 &25 &28 
&1.5 &2. \\
0518.8$-$6939 &  05~~18~~46.7 ~~$-$69~~39~~04 &0.0107$\pm$0.0027 &
100 &57 &11.7 &SNR \\  
~~~~~~~= CAL 23 \\
0519.8$-$6926 &  05~~19~~46.8 ~~$-$69~~25~~59 &0.0184$\pm$0.0032 &
 100 & 57 &7.0 & P1; SNR \\  
~~~~~~~= CAL 27 &= W520.1$-$6929 \\
0536.0$-$6736 &  05~~36~~00.1 ~~$-$67~~35~~58 &0.0046$\pm$0.0014 & 
 50 &61 & 18.0 & P1; SNR \\ 
~~~~~~~= CAL 60 &= W536.2$-$6737 \\ 
0537.8$-$6910 &  05~~37~~46.3 ~~$-$69~~10~~19 &0.0350$\pm$0.0040 & 
30 &63 &10.4 & 3.; P2; SNR; pulsar \\ 
~~~~~~~= CAL 67 &= W538.1$-$6912 \\ 
0538.8$-$6910 & 05~~38~~45.0 ~~$-$69~~10~~10 &0.0456$\pm$0.0083 &
 300 &62 & 14.1 & very diffuse; SNR? \\  
~~~~~~~= CAL 73 &= W539.2$-$6911 \\
0540.2$-$6920 & 05~~40~~12.0 ~~$-$69~~19~~52 &0.2158$\pm$0.0099 & 
 50 &62 & 11.0 & 4.; var; pulsar; \\ 
~~~~~~~= CAL 79 & & & & & & ~~~ PSR 0540$-$69 \\ 
\hline
\\
\end{tabular}
%\end{flushleft}

$^a$~SMC sources (Wang \& Wu 1992);
CAL$=$LHG sources (Long et al.\ 1981); W sources (Wang et al.\ 1991) \\
$^b$~From Table 1 \\
$^c$~Angular distance of source from center of field \\
$^d$~P1 or P2 indicates source was measured in Paper I or II in
another $ROSAT$ image \\
{\bf Notes:} \\
1. Probably a point source; X-ray pulsar with P=59.07 s \\
2.~~Obs.\#28 is from Paper~II \\
3.~~In Obs.\#62 at 0.05 ct s$^{-1}$ with 
100$^{\prime\prime}$ aperture and 14.0$^{\prime}$ from center \\
4.~~Also detected in Obs.\#63 at 19.4$^{\prime}$ from center, 
so count rate unreliable \\

\end{table}
\end{scriptsize}

\clearpage

\begin{scriptsize}

\begin{table}
\caption[]{SMC \& LMC X-ray Sources Not Detected by {\it ROSAT}}
\begin{tabular}{llcccc}
\hline
\hline
Source & ~~~~~~~~~Position$^a$ & Upper limit$^b$ & Obs.$^c$ &~Expected 
&Offset$^e$ \\ 
&~~~~R.A. ~~~~~~~~~~Dec. & 3$\sigma$ &\# & Count Rate$^d$ \\ 
 & ~~~~~~~~(J2000.0) & (cts s$^{-1}$) & &(cts s$^{-1}$) & ( $\prime$ ) \\
\hline
SMC 38 & 00$^h$56$^m$03$^s$ 
~$-$72$^\circ$~21$^\prime$~32$^{\prime\prime}$ &$<$0.002 &53 &0.002 &13.4 \\
SMC 39 & 00~~57~~22 ~~$-$72~~25~~46 &$<$0.006 & 53 &0.002 & 8.0  \\
SMC 40 & 00~~57~~32 ~~$-$72~~13~~17 &$<$0.003 & 53 &0.005 & 11.8  \\
SMC 42 & 00~~58~~16 ~~$-$72~~17~~39 &$<$0.008 & 53 &0.006 & 6.1  \\
RX~J0058.6$-$7146 & 00~~58~~36 ~~$-$71~~46~~02 &$<$0.002 & 52 &... & 10.0 \\
\\

SMC 47 & 01~~01~~44 ~~$-$72~~24~~20 &$<$0.003 &53 &0.006 &12.5 \\
CAL 25=W519.8$-$6921 & 05~~19~~27 ~~$-$69~~18~~00 &$<$0.011 &57 &0.007 &14.9 \\
RX~J0520.5$-$6932 & 05~~20~~29 ~~$-$69~~31~~52 &$<$0.003 &57 &0.009 &0.2  \\
W521.5$-$6921 & 05~~21~~10 ~~$-$69~~18~~31 & $<$0.005 &57 &0.012 &13.7 \\
W524.2$-$6937 & 05~~23~~47 ~~$-$69~~34~~39 & $<$0.005 &57 &0.055 &17.4\\
\\

RX~J0526.3$-$6944 & 05~~26~~18 ~~$-$69~~44~~01 &$<$0.002 &58 &0.008 &12.6  \\
RX~J0527.8$-$6954 & 05~~27~~48 ~~$-$69~~54~~16 &$<$0.002 &58 &0.008 &0.3  \\
W528.5$-$6942 &05~~28~~06 ~~$-$69~~40~~08 &$<$0.007 &58 &0.012 &14.0 \\
RX~J0532.0$-$6920 & 05~~32~~00 ~~$-$69~~19~~51 &$<$0.004 &59 &0.005 &0.1 \\
W533.2$-$7142 & 05~~32~~21 ~~$-$71~~40~~00 &$<$0.003 &60 &0.007 &0.4  \\
\\

W533.1$-$6928 & 05~~32~~43 ~~$-$69~~26~~28 &$<$0.005 &59 &0.004 &7.7  \\
CAL 55 & 05~~34~~57 ~~$-$69~~20~~13 &$<$0.006 &59 &0.022 &15.7  \\
CAL 55 & 05~~34~~57 ~~$-$69~~20~~13 &$<$0.004 &62 &0.022 &17.9  \\
W536.2$-$6921 & 05~~35~~48 ~~$-$69~~18~~43 &$<$0.003 &62 &0.008 &14.2  \\
RX~J0535.8$-$6732 & 05~~35~~48 ~~$-$67~~32~~25 &$<$0.003 &61 &0.002 &14.7 \\
\\

RX~J0536.3$-$6718 & 05~~36~~23 ~~$-$67~~17~~56 &$<$0.002 &61 &0.006 &0.6  \\
CAL 62=W536.8$-$6914 & 05~~36~~29 ~~$-$69~~11~~55 &$<$0.013 &62 &0.021 &15.4 \\
CAL 62=W536.8$-$6914 & 05~~36~~29 ~~$-$69~~11~~55 &$<$0.011 &63 &0.021 &16.3 \\
CAL 63 & 05~~36~~59 ~~$-$69~~31~~15 &$<$0.005 &62 &0.035 &10.0  \\
W538.0$-$6904 & 05~~37~~40 ~~$-$69~~02~~45 &$<$0.006 &63 &0.041 &6.6  \\
\hline

\end{tabular}

\end{table}
 
\clearpage

\begin{table}

{\normalsize Table 4 (cont): X-ray Sources Not Detected by {\it ROSAT}}\\
\begin{tabular}{llcccc}
\hline
\hline
Source & ~~~~~~~~~Position$^a$ & Upper limit$^b$ & Obs.$^c$ &~Expected 
&Offset$^e$ \\ 
&~~~~R.A. ~~~~~~~~~~Dec. & 3$\sigma$ &\# & Count Rate$^d$ \\ 
 & ~~~~~~~~(J2000.0) & (cts s$^{-1}$) & &(cts s$^{-1}$) & ( $\prime$ ) \\
\hline

CAL 68=W538.4$-$6907 & 05$^h$38$^m$01$^s$~ 
$-$69$^{\circ}$05$^{\prime}$43$^{\prime\prime}$ &$<$0.010 &63 &0.036 &6.1 \\
W538.9$-$6914 & 05~~38~~32 ~$-$69~~12~~27 &$<$0.012 &62 &0.018 &11.6  \\
W538.9$-$6914 & 05~~38~~32 ~$-$69~~12~~27 &$<$0.009 &63 &0.018 &10.8  \\
CAL 72=W539.1$-$6908 & 05~~38~~46 ~$-$69~~05~~58 &$<$0.020 &63 &0.065 &4.2 \\
CAL 74=W539.2$-$6911 & 05~~38~~52 ~$-$69~~02~~11 &$<$0.016 &63 &0.045 &0.4 \\
\\

W539.4$-$6900 & 05~~39~~05 ~~$-$68~~57~~59 &$<$0.018 &63 &0.018 &4.0  \\
W539.7$-$6858 & 05~~39~~20 ~~$-$68~~56~~14 &$<$0.012 &63 &0.003 &6.1  \\
W539.8$-$6907 & 05~~39~~27 ~~$-$69~~05~~27 &$<$0.015 &63 &0.012 &4.8  \\
W540.3$-$6909 & 05~~39~~57 ~~$-$69~~07~~45 &$<$0.010 &63 &0.019 &8.2  \\
CAL 77=W540.5$-$6927 & 05~~40~~04 ~~$-$69~~25~~37 &$<$0.006 &62 &0.022 &9.5 \\
\\

W541.0$-$6859 & 05~~40~~43 ~~$-$68~~57~~16 &$<$0.015 &63 &0.005 &10.9  \\
CAL 80=W541.8$-$6906 & 05~~41~~30 ~~$-$69~~04~~46 &$<$0.008 &63 &0.013 &14.3 \\
RX~J0547.9$-$6823 & 05~~47~~54 ~~$-$68~~22~~47 &$<$0.001 &65 &0.006 &0.2  \\
RX~J0554.2$-$6752 & 05~~54~~13 ~~$-$67~~52~~19 &$<$0.002 &68 &0.002 &11.5  \\
CAL 96   & 05~~54~~59 ~~$-$67~~48~~24 &$<$0.003 &68 &0.022 &6.0  \\
\hline
\\
\end{tabular}

$^a${\it Einstein} positions for CAL, W, \& SMC sources; {\it ROSAT} 
positions for RX~J sources \\
$^b$ All values background subtracted count rates using 
100$^{\prime\prime}$ diameter aperture \\
$^c$ From Table 1 \\
$^d$ Based on previous {\it ROSAT} or {\it Einstein} count rate \\
$^e$ Angular distance source expected to be from field center \\

\end{table}

\end{scriptsize}

\clearpage

\begin{scriptsize}

\begin{table}
\caption[]{Optically Identified SMC \& LMC {\it ROSAT} X-ray Sources}
%\begin{flushleft}
\begin{tabular}{llcccl}
\hline
\hline
Source  &~~~~~ Optical Position$^a$ & $V$ & Spectral & Finding & Remarks \& 
Notes$^c$ \\
&~~~~R.A. ~~~~~~~~~~~Dec.  && Type & Chart$^b$ \\
RX~J  & ~~~~~~~~ (J2000.0)  \\
\hline
0005.3$-$7427 & $00^h05^m20.4^s$ 
$-74^{\circ}26^{\prime}40.0^{\prime\prime}$ & 16.3 & AGN & Cr & 
resolved gal; z=0.1316 \\
 ~~~= SMC 1 & & & ~ \\
0033.3$-$6915 & 00~33~20.9 ~ $-$69~15~14.9 & 15.5 & galaxy cluster & Cr & 
A2789; z=0.0975; mag, posn, and \\
 ~~~= SMC 70 &&& ~~  && ~~~ redshift for central cD gal  \\
0051.9$-$7311 & 00~51~52.1 ~ $-$73~10~34.4 & 14.4 & Be pec & P2 &1.; Star 1 \\
 ~~~= SMC 25 \\
0052.9$-$7158 & 00~52~55.2 ~ $-$71~58~06.7 & 15.5$_{var}$ & Be star & P2 
& 1.; Star 1; var X-rays \\
 ~~~= SMC 32 \\
0058.2$-$7231 & 00~58~12.6 ~ $-$72~30~50.2  & 14.9 & Be star & P3 \\
0058.6$-$7136 & 00~58~37.4 ~ $-$71~35~49.2 & 16.6 & pl neb & P2 
& 1.; N67; `Star' 2; \\
 ~~~= SMC 43 &= 1E0056.8$-$7154 & && & ~~~ supersoft source  \\
0059.4$-$7210  & 00~59~11.7 ~ $-$72~10~14 & - - & SNR & - -  
&N66=0057$-$724=NGC346 \\
 ~~~= SMC 44  &&&&& ~~~= DEM S103 \\
0100.7$-$7212 & 01~00~43.0 ~ $-$72~11~42.9 & 13.5 & G star & P3 \\
 ~~~= SMC 45 \\
0117.5$-$7312 & 01~17~28.6 ~ $-$73~11~42.2  & 13.9 & G star & P3 & \\
0136.4$-$7105 & 01~36~25.3 ~ $-$71~05~24.9 & 19.1 & AGN & Cr & z=0.4598 \\
 ~~~= SMC 68 \\
0454.2$-$6643  & 04~54~11.0 ~ $-$66~43~16.6 & 18.2 & AGN & Cr & z=0.2279 \\
0510.8$-$6845 & 05~10~51.3 ~ $-$68~45~19.6 & 13.5: & K0~V & P2 & 1.; Star 
13; i.d. not certain \\
 ~~~= CAL 17 \\
0516.0$-$6916 & 05~16~00.1 ~ $-$69~16~08.5 & 15.0 & Be star & P2 & 1.; 
Star 2 \\
0518.8$-$6939 & 05~18~44.2 ~ $-$69~39~12 & - - &SNR & - - &N120=0519$-$697 \\
 ~~~= CAL 23 \\
0519.8$-$6926 & 05~19~45 ~ $-$69~25~59 & - -  & SNR & - -& 1.; 
SNR 0520$-$694 \\
 ~~~= CAL 27  \\
0531.5$-$7130 & 05~31~31.8 ~ $-$71~29~45.8 & 19.2  & AGN & P1 & 1.; `Star' 1; 
z=0.2214 \\
 ~~~= CAL 46  \\
0533.6$-$6733 & 05~33~36.7 ~ $-$67~32~43.9 & 14.0 & K star & P2 & 1.; 
Star 1  \\
0534.0$-$7145 & 05~33~59.1 ~ $-$71~45~22.0 & 13.8 & galaxy & Cr
& z=0.0242; Up 053448$-$7147.3; \\
&&&&& ~~~ X-ray source in disk of gal \\
0534.8$-$6739 & 05~34~44.7 ~ $-$67~38~54.3 & 17.8 & AGN & P2 & 1.; `Star' 2;
z=0.072 \\
 ~~~= W534.9$-$6741 \\
0536.0$-$6736 & 05~36~00.1 ~ $-$67~35~58  & - - & SNR & P1 & 
1.; DEM241=0536$-$676;  \\
 ~~~= CAL 60 &&&&& ~~~ (see note in Paper I)  \\

\hline

\end{tabular}

\end{table}
 
\clearpage

\begin{table}

{\normalsize Table 5 (cont): Optically Identified Sources} \\
\begin{tabular}{llcccl}
\\
\hline
\hline
Source  &~~~~~ Optical Position$^a$ & $V$ & Spectral & Finding & Remarks \& 
Notes$^c$ \\
&~~~~R.A. ~~~~~~~~~~~Dec.  && Type & Chart$^b$ \\
RX~J  & ~~~~~~~~ (J2000.0)  \\
\hline
0537.8$-$6910 & $05^h37^m51.8^s$ 
$-69^{\circ}10^{\prime}22^{\prime\prime}$ & - - & SNR  & - - & 1.; 
2.; N157B=DEM263=30 Dor B \\
 ~~~= CAL 67 &&&&& ~~~ =0538$-$691; contains 16 ms pulsar \\
0538.3$-$6924 & 05~38~16.3 ~ $-$69~23~31.3 & 12.0 & dMe & P2 & Star 1 \\
 ~~~= CAL 69 \\
0538.6$-$6853 & 05~38~34.6 ~ $-$68~53~06.8 & 10.2 & G2 V & P2 & RS CVn \\
 ~~~= CAL 71 \\
0538.8$-$6910 & 05~38~45:~ ~ $-$69~10~10:  & - - & SNR? & - - & 3.; N157A?\\
 ~~~= CAL 73 \\
0539.7$-$6945 & 05~39~39.2 ~ $-$69~44~36.5 & 14.8 & O7 III$_{pec}$ & P1 
&1.; 4.; var; Star 32; LMC X-1 \\
~~~= CAL 78 &&&& & \\
0540.2$-$6920 & 05~40~11.3 ~ $-$69~19~54.5 & 17.5  & pulsar & P3 & 5.; 
PSR 0540$-$69; \\
 ~~~= CAL 79  &&&&& ~~~ in SNR 0540$-$693 \\
0546.3$-$6835 & 05~46~15.0 ~ $-$68~35~23.6 & 19.2$_{var}$ 
& em star & P1 &1.; Star 2; dble H, He~I em \\
 ~~~= CAL 86? \\
0546.8$-$7109 & 05~46~46.4 ~ $-$71~08~52.6 & 19.5$_{var}$ & pec em & Sch & 
eclipsing supersoft binary  \\
 ~~~= CAL 87 \\
0549.8$-$7150 & 05~49~46.4 ~ $-$71~49~35.2 & 13.5 & dMe & P3 & 6.; H, 
Ca~II em \\
0550.5$-$7110 & 05~50~31.0 ~ $-$71~09~56.8 & 19.9$_{var}$ & AGN & Cr
& `Star' 2; z=0.4429 \\
\hline
\\
\end{tabular}

$^a$ Stars and AGN measured from CCD frames; SNR positions from 
Mathewson et al.\ 1983, 1985 or Davies et al.\ 1976 \\
$^b$ P1= Schmidtke et al.\ 1994, P2 = Cowley et al.\ 1997, P3 = this 
paper, Cr = Crampton et al.\ 1997; Sch = Schmidtke et al.\ 1993 \\
$^c$ Prefix `N' from Henize 1956; DEM = Davies et al.\ 1976 \\

{\bf Notes:}\\
1. X-ray position given in Papers I or II, Tables 2 or 3 \\
2. Pulsar found by Marshall et al.\ 1998 \\
3. X-ray source very large, irregular, and obviously associated with hot
gas in 30 Dor region; position is X-ray centroid; \\ 
 ~~~ may be associated with N157A; see deep X-ray image by Wang 1995 \\
4. Identification given by Cowley et al.\ 1995 \\
5. Position and magnitude from our CCD frames; integrated mag of 
plerion within 12$^{\prime\prime}$ aperture; \\
 ~~~ also see Caraveo et al.\ 1992 and Shearer et al.\ 1994 \\
6. Schmidtke \& Cowley 1996; Charles et al.\ 1996 \\
%\end{flushleft}
\end{table}
\end{scriptsize}

\clearpage

\begin{scriptsize}

\begin{table}
\caption[]{Summary of Identified SMC \& LMC Members in This Survey$^a$}
%\begin{flushleft}
\begin{tabular}{lllcl}
\hline
\hline
Source  &~~~~~ Optical Position & $V$ & Spectral & Remarks \& 
Notes \\
&~~~~R.A. ~~~~~~~~~~~Dec.  && Type  \\
RX~J  & ~~~~~~~~ (J2000.0)  \\
\hline
0037.3$-$7214  & 00~37~19.8 ~ $-$72~14~13.2 & 20.4$_{var}$ & pec em & 
supersoft X-ray binary  \\
~~~= SMC 13 & = 1E 0035.4$-$7230 \\
0048.3$-$7332 & 00~48~20.8 ~  $-$73~31~53.2 & 15.0$_{var}$ & symbiotic &
in cluster NGC 269 \\
0050.8$-$7316  & 00~50~44.8 ~ $-$73~16~05.8 & 15.4$_{var}$ &Be & \\
0051.9$-$7311 & 00~51~52.1 ~ $-$73~10~34.4 & 14.4 & Be pec & \\
 ~~~= SMC 25 \\
0052.9$-$7158 & 00~52~55.2 ~ $-$71~58~06.7 & 15.5$_{var}$ & Be star  
& X-rays variable \\
 ~~~= SMC 32 \\
0058.2$-$7231 & 00~58~12.6 ~ $-$72~30~50.2  & 14.9 & Be star  \\
0058.6$-$7136 & 00~58~37.4 ~ $-$71~35~49.2 & 16.6 & pl neb & N67; 
supersoft source \\
~~~= SMC 43 & = 1E0056.8$-$7154 \\
0516.0$-$6916 & 05~16~00.1 ~ $-$69~16~08.5 & 15.0 & Be star & \\
0117.1$-$7327 & 01~17~05.1 ~ $-$73~26~35.8 & 13.3$_{var}$ & B0 Ib & 
SMC X-1; MXRB  \\
~~~= SMC 63 & = 1E0114.5$-$7342  \\
0501.6$-$7034 & 05~01~23.7 ~ $-$70~33~33.8 & 14.5$_{var}$ & B0e & HV2289? \\
~~~= CAL 9\\
0502.9$-$6626 & 05~02~51.7 ~ $-$66~26~26.5 & 14.2$_{var}$ & B0e & 1.; 4.1-s 
X-ray pulsar \\
~~~= CAL E &&&& \\
0520.5$-$6932 & 05~20~29.8 ~ $-$69~31~54.8 &14.4$_{var}$  & O8e & \\
0539.7$-$6945 & 05~39~39.2 ~ $-$69~44~36.5 & 14.8$_{var}$ & O7 III$_{pec}$  
& LMC X-1 \\
~~~= CAL 78  \\
0540.2$-$6920 & 05~40~11.3 ~ $-$69~19~54.5 & 17.5$_{var}$  & pulsar & 
PSR 0540$-$69; \\
 ~~~= CAL 79  &&&& ~~~ in SNR 0540$-$693 \\
0543.6$-$6822 & 05~43~34.2 ~ $-$68~22~21.7 &17.3$_{var}$ &pec em 
& supersoft X-ray binary \\
~~~= CAL 83\\
0546.8$-$7109 & 05~46~46.4 ~ $-$71~08~52.6 & 19.5$_{var}$ & pec em  & 
eclipsing supersoft X-ray binary \\
 ~~~= CAL 87 \\
\hline
\\

\end{tabular}

$^a$Not including supernova remnants. \\

{\bf Notes:} \\
1. ~~See Schmidtke et al.\ 1995 \\

%\end{flushleft}
\end{table}
\end{scriptsize}

\clearpage

\begin{figure}
\caption{Observed $ROSAT$-HRI count rate for sources from this paper,
Paper~I, and Paper~II versus expected $ROSAT$ count rate, based on
observed {\it Einstein}-IPC fluxes.  The $+$ indicates sources which were
detected by both instruments.  Horizontal bars show the expected range of
HRI counts for assumed spectral models, while vertical bars are the
$\pm1\sigma$ errors for the observed $ROSAT$-HRI count rates.  A high
percentage of the detected sources appear to be variable.  Undetected
sources significantly to the right of the dashed line must be variable
(see discussion in text).  } 
\end{figure}

\begin{figure} 
\caption{Finding charts for X-ray point sources from CCD $V$ frames
obtained at CTIO.  The $+$ marks the $ROSAT$-HRI position.  Numbered
candidates are discussed in the text.  If no numbers are given and the
optical identification is known, the counterpart is marked by two dashes.
The sources are: 
RX~J0054.9$-$7227 (SMC 35; possibly Star 1 which is a Be star), 
RX~J0058.2$-$7231 (Be star), 
RX~J0100.7$-$7212 (SMC 45; G star),
RX~J0117.5$-$7312 (G star),
RX~J0540.2$-$6920 (CAL 79; pulsar PSR 0540$-$69),
and RX~J0549.8$-$7150 (dMe with hydrogen and Ca~II emission).}

\end{figure} 


\clearpage

\begin{references}

\reference{} Alcock, C., et al. 1997, MNRAS, 291, L13 

\reference{} Buckley, D.A.H., Stevens, J.B., \& Coe, M.J. 1998, in ``The
Stellar Content of Local Group Galaxies", eds. P.A. Whitelock \& R.D.
Cannon, ASP Conf. Ser., in press 

\reference{} Caraveo, P.A., Bignami, G.F., Mereghetti, S., \& Mombelli, M.
1992, \apjl, 395, L103

\reference{} Charles, P.A., Southwell, K.A., \& O'Donoghue, D. 1996, IAU
Circ\ 6305 

\reference{} Chu, Y.-H., Snowden, S.L., \& Chang, T. 1996, The Rosat
Newsletter, 13, 26 

\reference{cow84} Cowley, A.P., Crampton, D., Hutchings, J.B.,
Helfand, D.J., Hamilton, T.T., Thorstensen, J.R., \& Charles, P.A. 1984,
ApJ, 286, 196 

\reference{} Cowley, A.P., Schmidtke, P.C., Anderson, A.L., \& McGrath, 
T.K. 1995, PASP, 107, 145

\reference{} Cowley, A.P., Schmidtke, P.C., Crampton, D., \& Hutchings, 
J.B. 1990, ApJ, 350, 288

\reference{} Cowley, A.P., Schmidtke, P.C., Hutchings, J.B, Crampton, D.,
\& McGrath, T.K. 1993, ApJ, 418, L63

\reference{} Cowley, A.P., Schmidtke, P.C., McGrath, T.K., Ponder, A.L., 
Fertig, M.R., Hutchings, J.B., \& Crampton, D. 1997, \pasp, 109, 21
(Paper~II)

\reference{} Cowley, A.P., Schmidtke, P.C., Taylor, V.A., McGrath, T.K.,
Hutchings, J.B., \& Crampton, D. 1998, in ``The Stellar Content of Local
Group Galaxies", eds. P.A. Whitelock \& R.D. Cannon, ASP Conf. Ser., in press 

\reference{cra85} Crampton, D., Cowley, A.P., Thompson, I.B., \&
Hutchings, J.B. 1985, AJ, 90, 43 

\reference{} Crampton, D., Gussie, G., Cowley, A.P., \& Schmidtke, P.C.
1997, \aj, 114, 2353

\reference{dav76} Davies, R.D., Elliott, K.H., \& Meaburn, J. 1976,
Mem.\ R.\ Ast.\ Soc., 81, 89

\reference{} Greiner, J., Schwarz, R., Hasinger, G., \& Orio, M. 1996,
Lecture Notes in Physics, 472, 145, ed. J. Greiner, (Springer-Verlag: Berlin) 

\reference{hen56} Henize, K.G. 1956, ApJS, 2, 315

\reference{} Hutchings, J.B., Crampton, D., Cowley, A.P., \& Schmidtke,  
P.C. 1998, ApJ, 502, 408

\reference{} Kahabka, P. \& van den Heuvel, E.P.J. 1997, AARA, 35, 69

\reference{kur93} Kuerster, M. 1993, {\it ROSAT} Status Report \#67

\reference{lan92} Landolt, A.U. 1992, \aj, 104, 340

\reference{las90} Lasker, B.M., Sturch, C.R., McLean, B.J., 
Russell, J.L., Jenkner, H., \& Shara, M.M. 1990, \aj, 99, 2019

\reference{lon81} Long, K.S., Helfand, D.J., \& Grabelsky, D.A. 1981, 
ApJ, 248, 925

\reference{} Marshall, F.E., Gotthelf, E.V., Zhang, W., Middleditch, J., 
\& Wang, Q.D. 1998, ApJ, 499, L179

\reference{} Marshall, F.E. \& Lochner, J.C. 1998, IAU Circ 6818

\reference{mat83} Mathewson, D.S., Ford, V.L., Dopita, M.A., Tuohy, I.R., 
Long, K.S., \& Helfand, D.J. 1983, ApJS, 51, 345

\reference{mat85} Mathewson, D.S., Ford, V.L., Tuohy, I.R., Mills, B.Y., 
Turtle, A.J., \& Helfand, D.J. 1985, ApJS, 58, 197

\reference{} Schmidtke, P.C. \& Cowley, A.P. 1996, Lecture Notes in
Physics, 472, 123, ed. J. Greiner, (Springer-Verlag: Berlin) 
 
\reference{sch94} Schmidtke, P.C., Cowley, A.P., Frattare, L.M., McGrath,
T.K., Hutchings, J.B., \& Crampton, D. 1994, \pasp, 106, 843 (Paper~I)

\reference{} Schmidtke, P.C., Cowley, A.P., McGrath, T.K., \& Anderson, 
A.L. 1995, PASP, 107, 450

\reference{} Schmidtke, P.C., McGrath, T.K., Cowley, A.P., \& Frattare, L.M.
1993, \pasp, 105, 863

\reference{sew81} Seward, F.D. \& Mitchell, M. 1981, \apj, 243, 736

\reference{} Shearer, A., Redfern, M., Pedersen, H., Rowold, T., O'Kane, 
P., Butler, R., O'Byrne, C., \& Cullum, M. 1994, \apjl, 423, L51

\reference{} Stetson, P.B. 1987, \pasp, 99, 191

\reference{} Thomas, H.-C. 1996, private communication

\reference{}van Paradijs, J. 1995, in ``X-ray Binaries", ed. W.H.G. Lewin, 
J. van Paradijs, \& E.P.J. van den Heuvel, (Cambridge University Press: 
Cambridge), 536

\reference{wang91} Wang, Q. 1991, MNRAS, 252, 47P

\reference{} Wang, Q.D. 1995, ApJ, 453, 783

\reference{} Wang, Q.D. \& Gotthelf, E.V. 1998, ApJ, 494, 623

\reference{wan91} Wang, Q., Hamilton, T., Helfand, D.J., \& Wu, X. 1991,
ApJ, 374, 475 

\reference{wan92} Wang, Q. \& Wu, X. 1992, \apjs, 78, 391

\reference{} Ye, T., Turtle, A.G., \& Kennicutt, R.C. 1991, \mnras, 249, 722

\end{references}
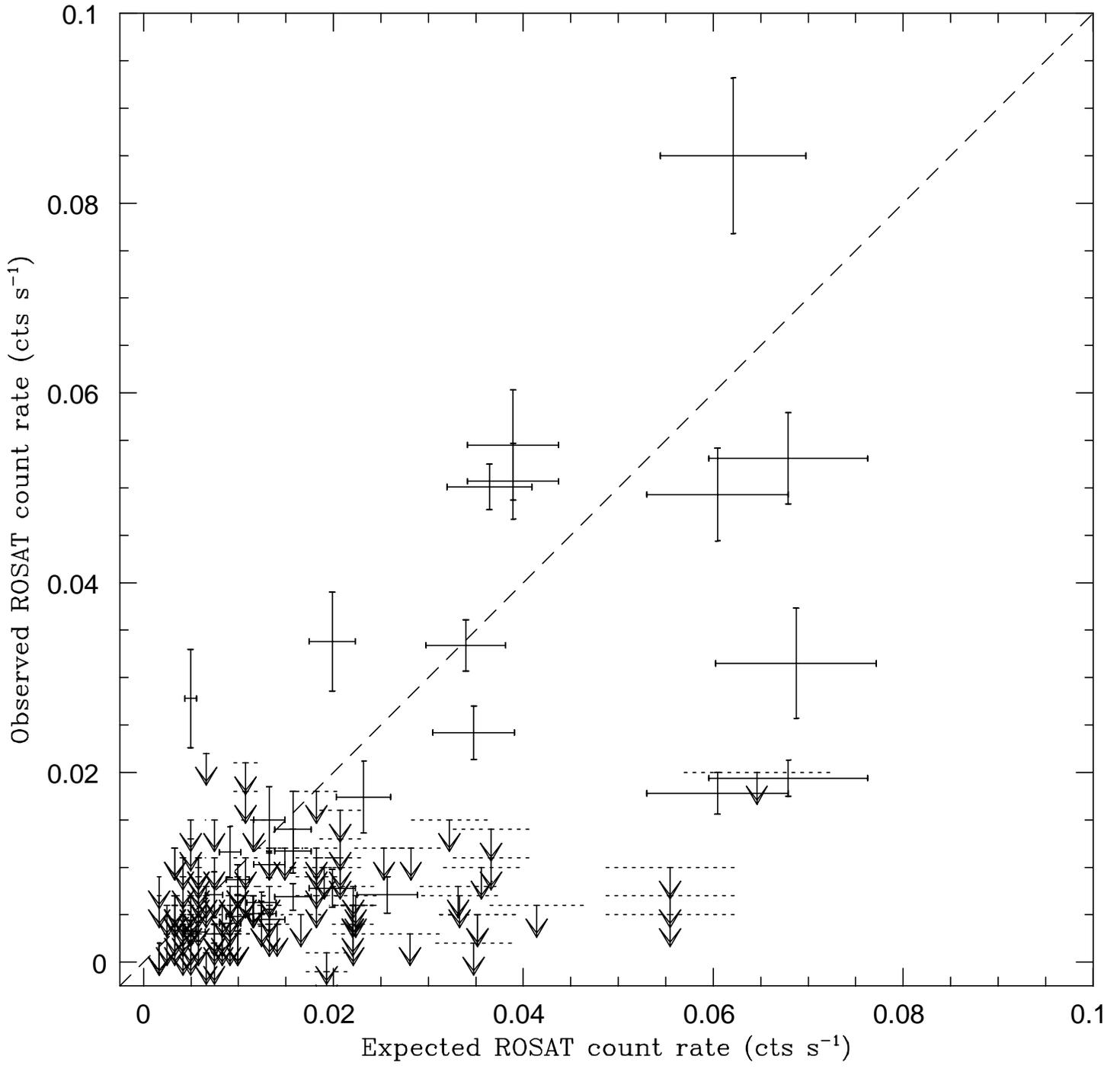
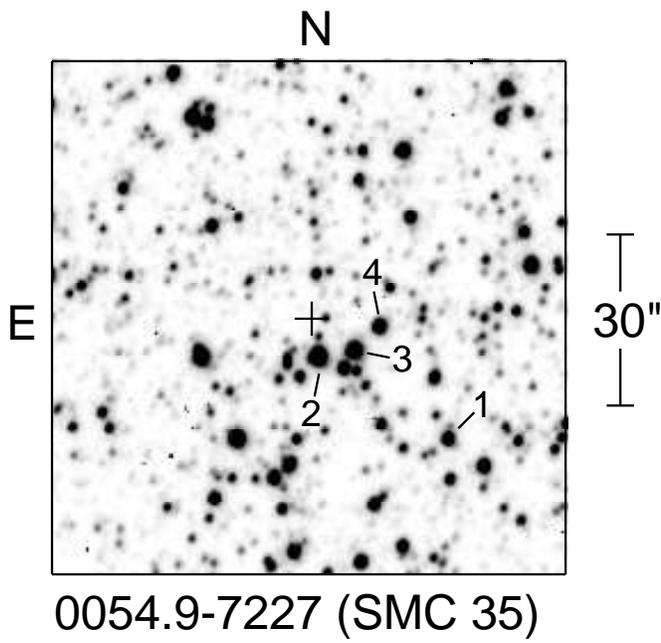
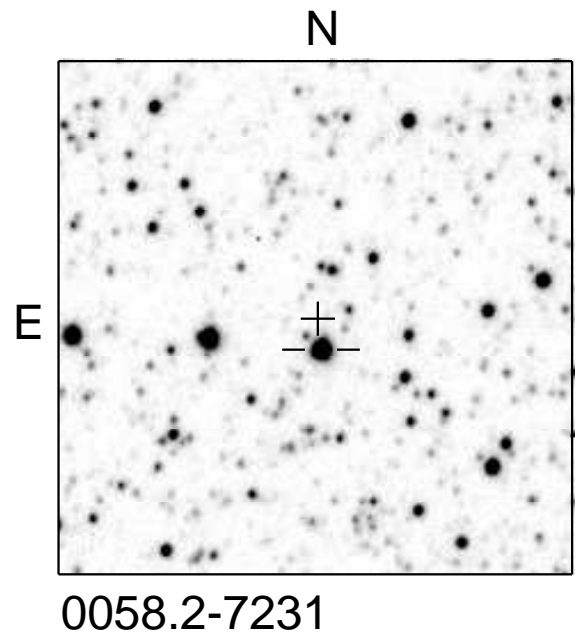
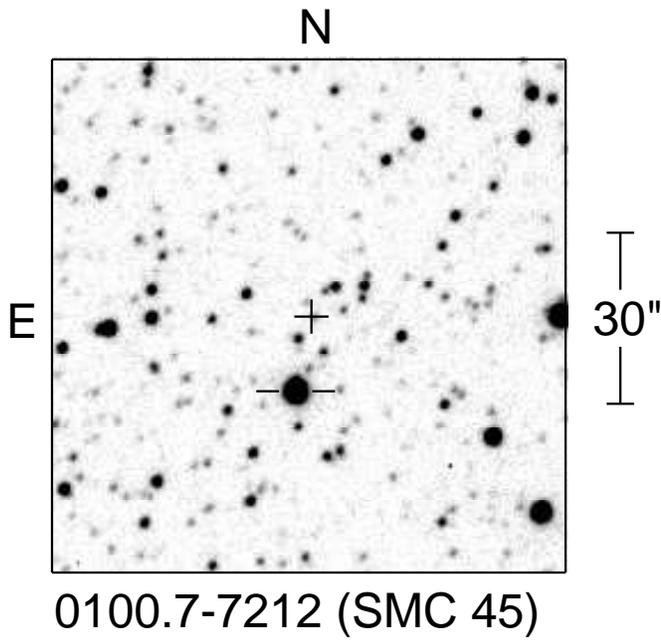
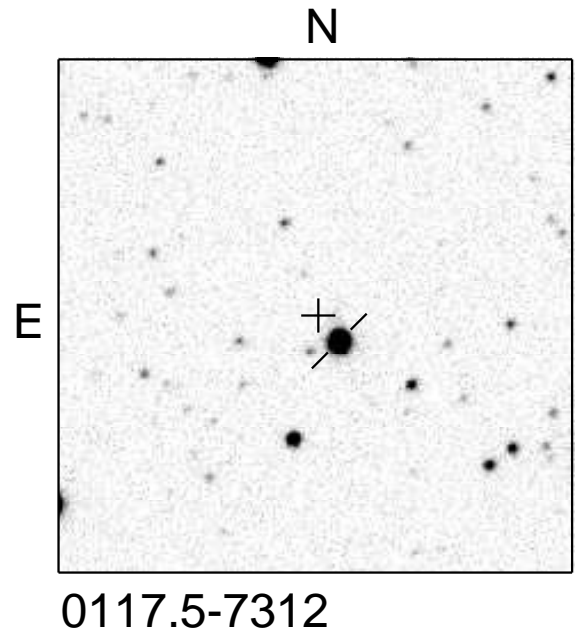
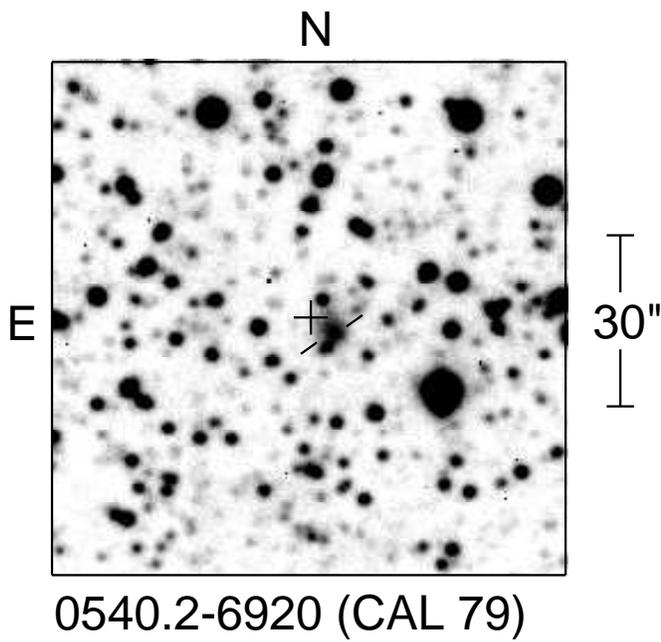
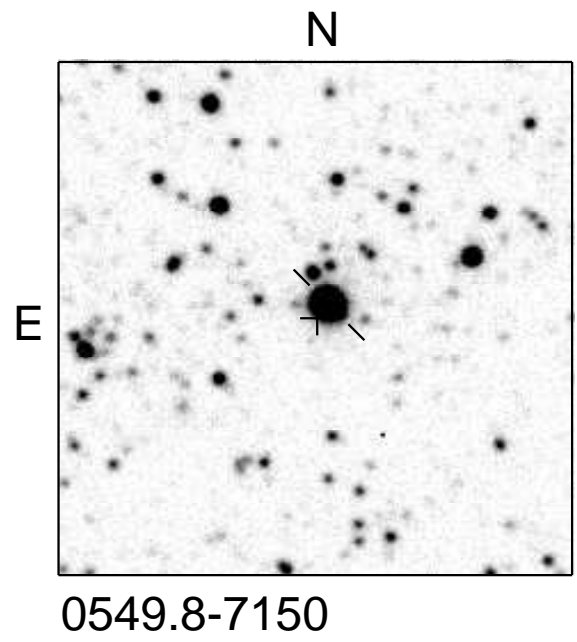
\end{document}